\documentclass[sigconf]{acmart}

\usepackage{booktabs} 
\usepackage{graphicx}
\usepackage[linesnumbered,ruled,noend]{algorithm2e}
\usepackage{multirow}
\usepackage{threeparttable}
\usepackage{tikz}
\usepackage{pgfplots}
\usepackage{pgfplotstable}
\usepackage{caption}
\usepackage{subcaption}

\newcommand{\nop}[1]{}%
\newcommand{\Paragraph}[1]{~\vspace*{-0.9\baselineskip}\\{\bf #1}}

\copyrightyear{2017}
\acmYear{2017}
\setcopyright{acmcopyright}
\acmConference{CIKM'17 }{November 6--10, 2017}{Singapore,
	Singapore}\acmPrice{15.00}\acmDOI{10.1145/3132847.3132957}
\acmISBN{978-1-4503-4918-5/17/11}

\begin{document}
\title{Keyword Search on RDF Graphs --- A Query Graph Assembly Approach}

 \author{
	{Shuo Han{${^1}$}, Lei Zou{${^1}$}, Jeffery Xu Yu{${^2}$}, Dongyan Zhao{${^1}$}} \\
	\fontsize{10}{\baselineskip}\selectfont\itshape $~^{1}$Peking University, China;\\
	\fontsize{10}{\baselineskip}\selectfont\itshape $~^{2}$ The Chinese University of Hong Kong, China; \\
	\fontsize{9}{\baselineskip}\selectfont\ttfamily\upshape $\{$hanshuo,zoulei,zhaody$\}$@pku.edu.cn, yu@se.cuhk.edu.hk\\
}

\begin{abstract}
Keyword search provides ordinary users an easy-to-use interface for querying RDF data. Given the input keywords, in this paper, we study how to assemble a query graph that is to represent user's query intention accurately and efficiently. Based on the input keywords, we first obtain the elementary query graph building blocks, such as entity/class vertices and predicate edges. Then, we formally define the \emph{query graph assembly (QGA)} problem. Unfortunately, we prove theoretically that QGA is a NP-complete problem. In order to solve that, we design some heuristic lower bounds and propose a bipartite graph matching-based best-first search algorithm. The algorithm's time complexity is $O(k^{2l} \cdot l^{3l})$, where $l$ is the number of the keywords and $k$ is a tunable parameter, i.e., the maximum number of candidate entity/class vertices and predicate edges allowed to match each keyword. Although QGA is intractable, both $l$ and $k$ are small in practice. Furthermore, the algorithm's time complexity does not depend on the RDF graph size, which guarantees the good scalability of our system in large RDF graphs. Experiments on DBpedia and Freebase confirm the superiority of our system on both effectiveness and efficiency. 
\end{abstract}

%
%
%


\maketitle

\section{Introduction} \label{sec:introduction}

RDF data is enjoying an increasing popularity as information extraction and archiving techniques advance. Many efforts such as DBpedia \cite{lehmann2015dbpedia} and Freebase \cite{bollacker2008freebase} have produced large scale RDF repositories. The use of RDF data has further gained popularity due to the launching of ``knowledge graph'' by Google and Linked Open Data project by W3C. One of fundamental issues in RDF data management is how to help end users obtain the desired information conveniently from RDF repositories, which has become a challenging problem and attracted lots of attentions in the database community . Although SPARQL is a standard query language to access RDF data, it is impractical for ordinary users to write SPARQL statements due to the complexity of SPARQL syntax and the lack of priori knowledge of RDF datasets.
Therefore, some efforts have been made to provide easy-to-use interfaces for non-professional users, such as natural language question answering (NL-QA for short) \cite{unger2012template,yahya2012natural,zou2014natural} and keyword search \cite{tran2009top,pound2012interpreting,le2014scalable} over RDF datasets. Ideally, the interfaces take natural language (NL) sentences (e.g. ``which scientist graduate from a university located in USA?'') or keyword queries(e.g. ``scientist graduate from university locate USA'') as end user's input, and retrieve relevant answers. Compared to NL sentences, keywords are more concise and flexible. A complete NL sentence can provide more semantic information than keywords through its syntactic structure. For example, \cite{yahya2012natural,zou2014natural} derive query graph structure by syntactic rules on dependency trees \cite{de2008stanford}, which is unavailable for keywords. Hence keyword search brings more technical challenges, such as disambiguation and query intention understanding. Due to the proliferation of Web search engines, the keyword based information retrieval mechanism enjoys widespread usage, especially in ``search box'' based applications. Therefore, this paper focuses on keyword search over RDF graphs. There are two technical challenges that should be addressed for a desirable keyword search system.  

\textbf{Effectiveness: understanding the query intention \emph{accurately}}. Although keywords are concise and flexible for end users, it is not an easy task for a system to understand the query intention behind that. Generally, there are two obstacles. First, it is the ambiguity of keywords. Given a keyword, we may have multiple ways to interpret the keyword. A system should figure out which interpretation is correct given the context of the keywords. The second obstacle is the ambiguity of query structures. Even if each keyword has been correctly interpreted, how to represent the whole query's intention is also a challenging task.
A desirable representation of query intention is to interpret the input keywords as a structural query (such as SPARQL) that can be evaluated over the underlying RDF dataset to retrieve the answers that are of interests to users. Therefore, we aim to design an effective interpreting mechanism to translate the input keywords into SPARQL statements.  
	
\textbf{Efficiency: scaling to large graphs \emph{efficiently}}. RDF datasets tend to be very large. For example, Freebase contains about 1.9 billion triples and DBpedia has more than 583 million triples (in our experiments). As an online application over RDF graphs, the keyword search system's efficiency is another criteria. The system's performance consists of two parts: one is to translate the input keywords as SPARQL statements and the other is to evaluate SPARQL. Note that the latter is not the focus of this paper, as it has been studied extensively \cite{zou2014gstore,shi16fast}. We concentrate on building an efficient keyword interpreting mechanism.  

\subsection{Our approach}
To address the effectiveness and efficiency challenges discussed above, we propose a \emph{query graph assembly} (QGA for short) problem to model the keyword search task in this paper. In brief, we firstly match each keyword term into a group of elementary \emph{query graph element} candidates (i.e. entity/class vertices and predicate edges), then assemble these graph elements into a query graph $Q$, which can express users' query intention accurately. 

For the effectiveness issue, our QGA solution integrates the keyword disambiguation and query graph generation into a uniform model (i.e., the Assembly Bipartite Graph model in Definition \ref{def:assemblygraph}) under a single objective function. The uniform model overcomes the error propagation in the separated solution (such as first disambiguating the input keywords and then determining the query structure). 

For the efficiency issue, it is worth noting that the complexity of the QGA problem is \emph{irrelevant to the underlying data graph size}. Although QGA is intractable (we prove the NP-completeness of QGA in Section \ref{sec:hardness}), we figure out the overall search space only depends on the number of keyword terms $l$ and a tunable parameter $k$, i.e., the maximum number of candidate entity/class vertices and predicate edges allowed to match each keyword term. Both $l$ and $k$ are small in practice. Consequently, the time complexity is independent to the scale of RDF graph $G$, which theoretically explains why our approach is much faster than the comparative work over large RDF graphs.

In a nutshell, our contributions are summarized as follows:

\begin{enumerate}
	\item We formulate the keyword search task as a \emph{query graph assembly} problem, and integrate the keyword disambiguation and query graph formulation into a uniform model. It achieves higher accuracy than the widely adopted small-size connected structure criteria, such as the group Steiner tree. 	
	\item We theoretically prove that QGA is NP-complete. To solve that, we propose a constraint-based bipartite graph matching solution, which is a practical efficient algorithm with bound-based pruning techniques. 
	\item We employ graph embedding technique to measure the goodness of query graph $Q$.
	\item We conduct extensive experiments on real RDF graphs
	to confirm the the effectiveness and efficiency of our approach.
\end{enumerate}

\section{Related Work} \label{sec:related_work}

\begin{figure} [t]
	\centering
	\scalebox{0.8}[0.9]
	{
		\resizebox{\linewidth}{!}
		{
			\includegraphics[scale=1.0]{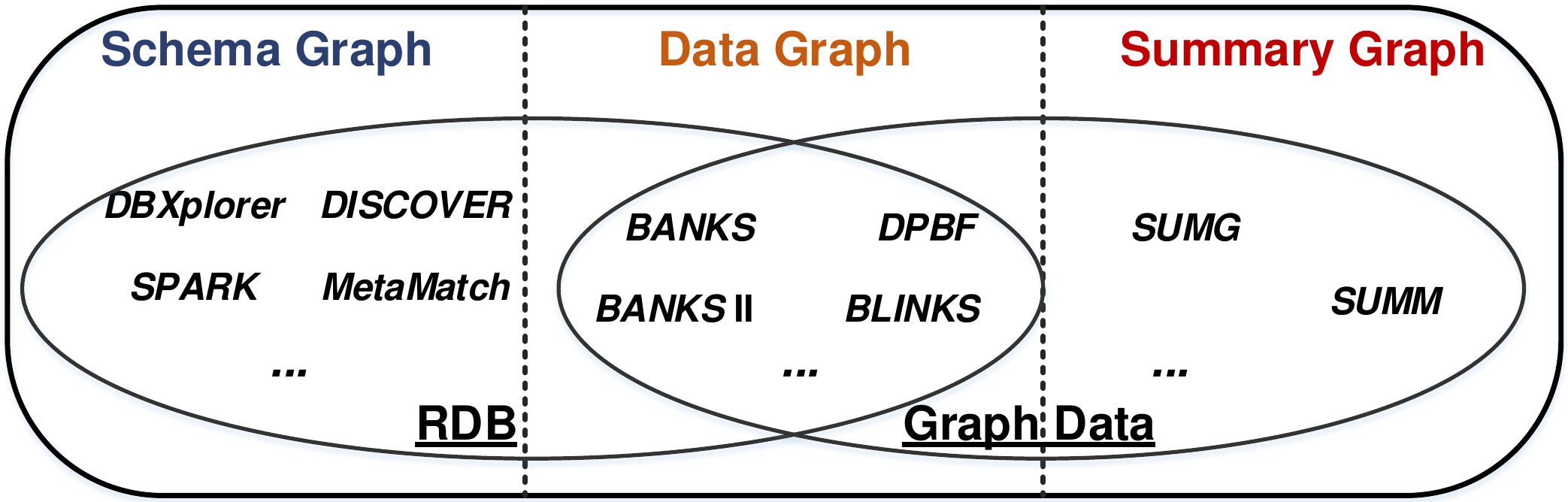}
		}
	}
	\caption{Related Work of Keyword Search Problem.}
	\label{fig:related_work_category}
	\vspace{-0.2in}
\end{figure}

There are many proposals dedicated to keyword search problem on relational database (RDB for short) and graph data. We summarize them into three categories as illustrated in Figure \ref{fig:related_work_category}. The first category exploits the ``schema'' of RDB, so we call them schema based approaches, including DBXplorer \cite{agrawal2002dbxplorer}, DISCOVER \cite{hristidis2002discover}, SPARK \cite{luo2007spark}, and so on. According to foreign/primary key relationships between tables, these approaches first build a ``schema graph''. Then based on the schema graph, they find join trees to infer the query pattern's structure, namely, SQL. 
To cope with the ambiguity of keywords, MetaMatch \cite{bergamaschi2011keyword} proposes a Hungarian bipartite matching algorithm as a joint disambiguation mechanism of mapping keyword tokens into records in RDB. However, MetaMatch still relies on the schema graph to infer SQL structure. As we know, RDF employs ``schema-less'' feature, thus it is impossible to rely on the underlying schema structure to generate query graphs for an RDF dataset.  

Due to the lack of schema in RDF repository, some existing work (such as SUMG \cite{tran2009top} and SUMM \cite{le2014scalable}) propose class based summarization techniques to generate ``summary graph''. Firstly they leverage the hierarchy of RDF class nodes to generate the summary graph $G_S$, which is much smaller than $G$. Then the graph exploration algorithms can be employed on $G_S$ to retrieve relevant answers. However, the summary graph in SUMG collapses all the entities of the same class into one summary vertex. Thus it loses detailed information about relations between entities of the same class, which may lead to incomplete or incorrect answers.

The last category is employing graph exploration algorithms directly on ``data graphs'', to find small size structures connecting all keywords. The data graph can be derived from RDB as well as RDF dataset. In RDB, each record is modeled as an vertex, and a foreign/primary key relationship between two records is regarded as an edge. In RDF context, subjects and objects are regarded as vertices, and an triple is an edge connecting two corresponding vertices. Therefore, the data graph based approaches can be used in both RDB and RDF context. Formally, given a data graph $G$, the keyword search problem is defined to find a minimum connected tree \cite{ding2007finding} or subgraph \cite{elbassuoni2011keyword} over $G$ that covers all the keyword terms by the tree/subgraph's nodes/edges. The problem definition is similar to the group Steiner tree (GST for short) problem, which is NP-complete. BANKS \cite{bhalotia2002keyword}, BANKS II \cite{kacholia2005bidirectional}, and BLINKS \cite{he2007blinks} are search based approximate algorithms, whereas DPBF \cite{ding2007finding} is a dynamic programming algorithm that can find the exact optimal result, with parameterized time complexity.

As illustrated in Figure \ref{fig:related_work_category}, both data graph based and summary graph based approaches can be used in RDF repository. There are two common drawbacks in these two categories. 

The first one is that they do not intend to \emph{understand} the input keywords. One common assumption in existing work is that the smaller of the result tree's size the more semantic it contains, which should be more interesting to users. For example, the widely adopted GST model as well as its variants, such as r-radius Steiner graph \cite{li2008ease} and multi-center induced graph \cite{qin2009querying}. However, this \emph{implicit} semantic representation of keywords is not good at result's accuracy, since they do not comprehend the query intention behind the keywords.

The second one is that their online search algorithm's complexity depends on the underlying graph size. Some graph exploration algorithms like BANKS, BANKS II, and BLINKS build distance matrix over $G$ to speed up exploration. The size of distance matrix becomes prohibitively expensive when handling large RDF graphs. DPBF's complexity is $O(3^l|V|+2^l(|V|log|V|+|E|))$, where $l$ is the number of keywords. Although $l$ is small, the algorithm's performance still depends on the scale of $G$. As graph scale increases, efficiency is still a challenging issue. The derived summary graph is smaller than the original data graph, but the size of summary graph still subjects to the \emph{structuredness} of data graph. Open-domain RDF datasets usually have complicated class hierarchy and heterogeneous relations, which lead to a high variance in their structuredness. Obviously, an ideal algorithm on large graph requires that its time complexity does not depend on the underlying data graph size. This is another motivation of our solution. 

There are some work based on the ``supervised learning'' technique to interpret the query intention of keywords into structured queries \cite{pound2012interpreting}. They require manually annotated query logs to train a semantic term classifier and interpreting templates offline. We believe that it is quite expensive (even impossible) to employ human efforts to annotate comprehensive training data when coping with large RDF datasets. 

NL-QA \cite{unger2012template,yahya2012natural,zou2014natural} is another related area. The difference is that NL-QA system takes NL sentences rather than keywords as input. As mentioned in Section \ref{sec:introduction}, keywords do not have syntax structure, so that we can not derive query graph structures from dependency trees \cite{de2008stanford} like \cite{yahya2012natural,zou2014natural}. Even so, our system is still competitive (ranked at top-3) with other NL-QA systems in the QALD-6 competition. 
\section{Problem Definition}\label{sec:problemdef}

\begin{figure} [t]
	\centering
	\scalebox{1.05}[1.15]
	{
		\resizebox{\linewidth}{!}
		{
			\includegraphics[scale=1.0]{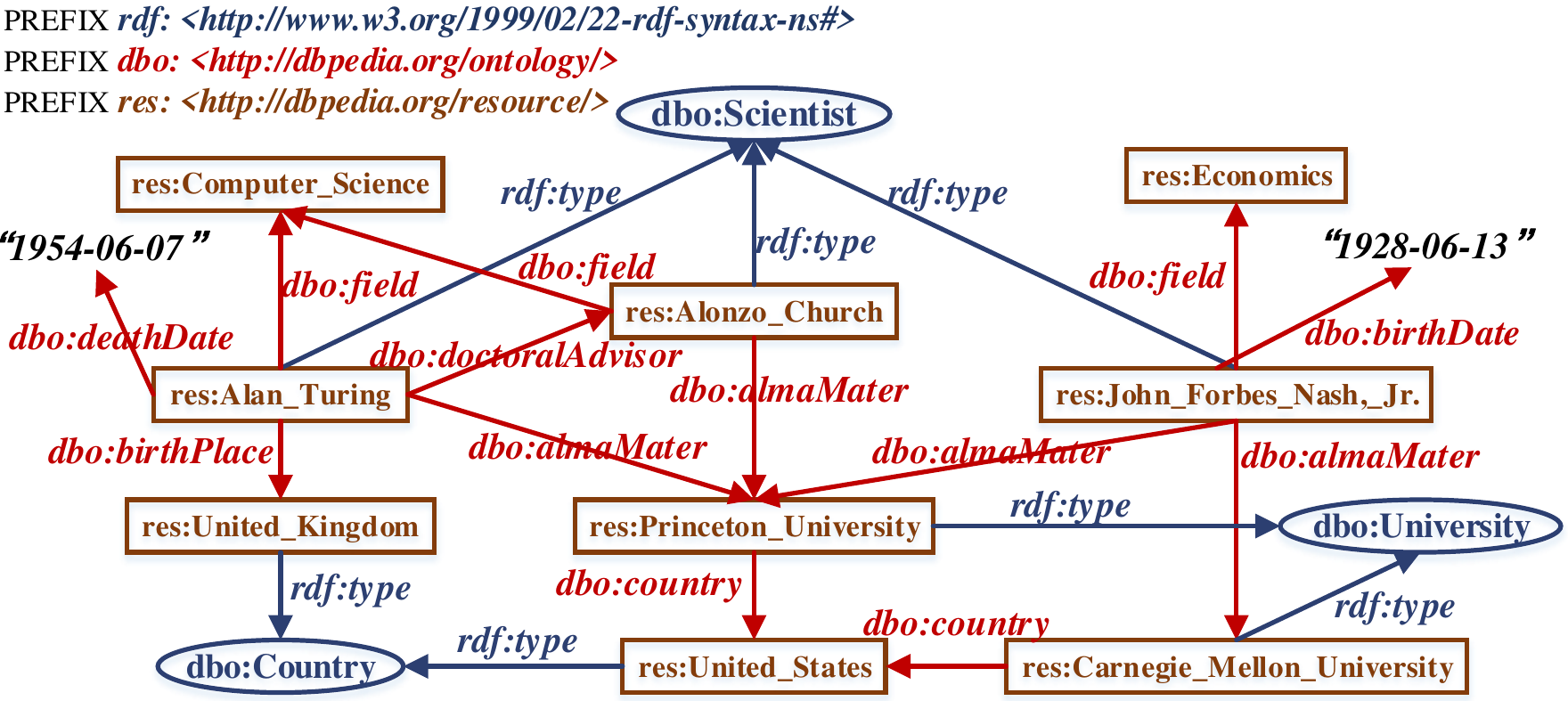}
		}
	}
	\vspace{-0.1in}
	\caption{A Sample of DBpedia RDF Graph.}
	\label{fig:rdfgraph}
	\vspace{-0.2in}
\end{figure}

In this section, we define our problem and review the terminologies used throughout this paper. As a de facto standard model of knowledge base, RDF represents the assertions by $\langle$subject, predicate, object$\rangle$ triples. An RDF dataset can be represented as a graph naturally, where subjects and objects are vertices and predicates denote directed edges between them. A running example of RDF graph is illustrated in Figure \ref{fig:rdfgraph}. Formally, we have the definition about RDF graph as follows.

\begin{definition} \label{def:rdfgraph}
	\textbf{ (RDF Graph) }
	An \textit{RDF graph} is denoted as $G(V, E)$, where
	$V$ is the set of entity and class vertices corresponding to subjects and objects of RDF triples,
	and $E$ is the set of directed relation edges with their labels corresponding to predicates of RDF triples.

\end{definition}

Note that RDF triple's object may be literal value, for example, $\langle$res:Alan\_Turing, dbo:deathDate, ``1954-06-07'' $\textasciicircum{}\textasciicircum{}$xms:date$\rangle$. We treat all literal values as entity vertices in RDF graph, and literal types (e.g. xms:date as for ``1954-06-07'') as class vertices. 

SPARQL is the standard structural query language of RDF, which can also be represented as a \emph{query graph} defined as follows.

\begin{definition} \textbf{ (Query Graph) }
	A \textit{query graph} is denoted as $Q(V_Q, E_Q)$, where
	$V_Q$ consists of entity vertices, class vertices, and vertex \emph{variables},
	and $E_Q$ consists of relation edges as well as edge \emph{variables}.
\end{definition}

In this paper, we study the keyword search problem over RDF graph. Given a keyword token sequence $RQ = \{k_1, k_2, ..., k_m\}$, our problem is to interpret $RQ$ as a \emph{query graph} $Q$. 

\section{System Overview} \label{sec:overview}

\begin{figure} [h]
	\centering
	\scalebox{1.0}[1.0]
	{
		\resizebox{\linewidth}{!}
		{
			\includegraphics[scale=0.5]{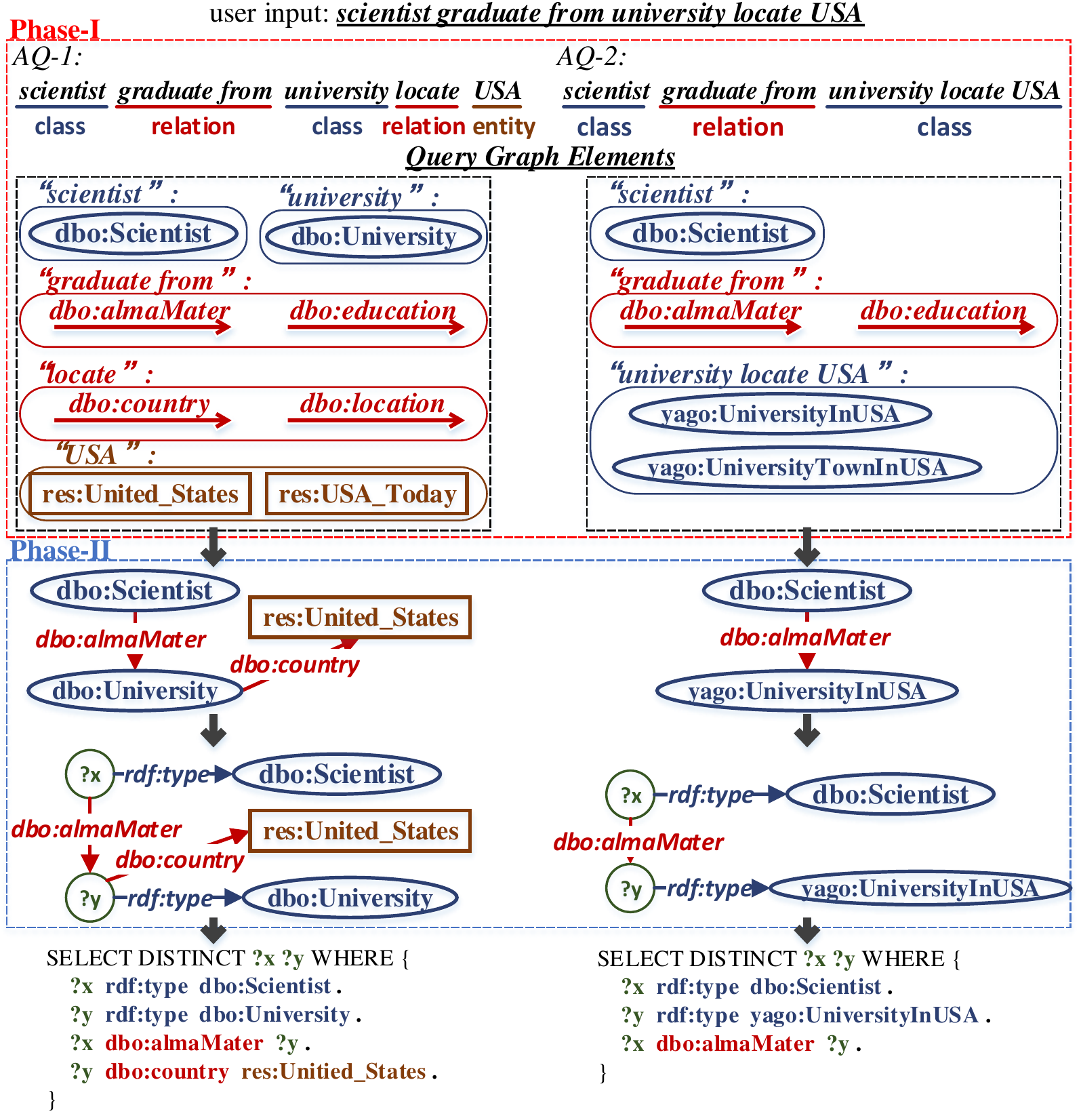}
		}
	}
	\vspace{-0.3in}
	\caption{An Overview of Our Approach.}
	\label{fig:overview}
	\vspace{-0.2in}
\end{figure}

In this section, we give an overview of our keyword search system. Our approach can be summarized as a two-phase framework, as illustrated in Figure \ref{fig:overview}. 

\subsection{Phase-I: Segmentation and Annotation} \label{sec:phase1}
The first phase is to segment the keyword token sequence $RQ = \{k_1, k_2, ..., k_m\}$ into several \emph{terms} and each term is annotated with one of the three characters $\{$entity, class, relation$\}$. The converted query is called \emph{annotated query}. Formally, we denote an \textit{annotated query} as $AQ= \{t_1:c_1, t_2:c_2, ..., t_l:c_l\}$, where each $t_i$ is a term and  $c_i \in \{$entity, class, relation$\}$. Note that the first phase (i.e., segmentation and annotation) is not the focus of this paper, as it has been studied extensively \cite{hua2015short,han2011generative, li2011faerie, nakashole2012patty,cai2013large}. We briefly describe the implementation of the first phase as follows. 

For each continuous subsequence $s$ in $RQ$, we check whether it could be matched to an entity, a class, or a relation of the RDF dataset, by employing the existing techniques of entity linking \cite{han2011generative,li2011faerie, ratinov2011local} and relation paraphrasing \cite{nakashole2012patty,cai2013large}. If $s$ is matched, we regard $s$ as a \emph{candidate term} $t_i$, and annotate it with the corresponding character (entity, class, or relation).
We may find that two candidate terms $t_i$ and $t_j$ \emph{overlap} with each other. We say $t_i$ overlaps with $t_j$ if and only if they have at least one common token. Obviously, if two terms overlap, they cannot occur at the same segmentation result. For example, ``university'' and ``university locate USA'' cannot occur in the same segmentation result. We build a \emph{candidate term graph} to describe the mutually exclusive relations: (1) each candidate term $t_i$ is represented a vertex; (2) there is an edge between $t_i$ and $t_j$ if and only of there is \emph{no} overlapping tokens between $t_i$ and $t_j$. Thus each maximal clique in the candidate term graph stands for a possible segmentation result. To obtain top-$N$ best $AQ$, we employ the maximal clique algorithm \cite{bron1973algorithm}, and adopt the pairwise metrics in \cite{hua2015short} to rank the segmentation result.
In out example, we get the top-2 $AQ$ as shown in Figure \ref{fig:overview}.

In the first phase, we have converted keyword token sequence $RQ$ into top-$N$ $AQ$ by some off the shelf techniques. Furthermore, these terms in $AQ$ have been matched to some elementary query graph building blocks (i.e., entity/class vertices and predicate edges). Specifically, if a term $t_i$ is annotated with ``entity'' or ``class'', it will be matched to candidate entity/class vertices in RDF graph.
If a term $t_i$ is annotated with ``relation'', it will be matched to a set of candidate predicates.

\vspace{-0.05in}
\begin{example}\label{example:aq}
	Given a keyword token sequence $RQ = \{$scientist, graduate, from, university, locate, USA$\}$, we obtain the annotated query $AQ=\{$``scientist'':class, ``graduate from'':relation, ``university'':class, ``locate'':relation, ``USA'':entity $\}$, where ``scientist'' is matched to $\{$dbo:Scientist$\}$, ``university'' is matched to $\{$dbo:University$\}$, ``USA'' is matched to two possible entities $\{$res:USA$\_$Today, res:United$\_$\\States$\}$ due to the ambiguity. Also, the relations ``graduate from'' and ``locate'' also match to two candidate predicates $\{$dbo:almaMater, dbo:education$\}$ and $\{$dbo:country, dbo:location$\}$
\end{example}

\vspace{-0.1in}
\subsection{Phase-II: Query Graph Assembly}
In the second phase, we concentrate on how to assemble a query graph $Q$ based on these elementary building blocks. Formally, the query graph assembly problem is defined as follows:

\vspace{-0.05in}
\begin{definition} \textbf{(Query Graph Assembly Problem)}\label{def:querygraphassembling}
	Given $n$ terms $t^{v}_i$ ($i=1,...,n$) annotated with ``entity'' or ``class'', and $m$ terms $t_j^{e}$ ($j=1,...,m$) annotated with ``relation'', each term $t^{v}_i$ is matched to a set $V_i$ of candidate entity/class vertices and each $t^{e}_j$ is matched to a set $E_j$ of candidate predicate edges. Let $\Upsilon=\{V_1, ..., V_n\}$ and $\Gamma=\{E_1, ..., E_m\}$. A valid assembly query graph is denoted as $Q (V_Q,E_Q)$, which satisfies the following constraints:
	\begin{enumerate}
		\item $|V_Q|=n$, and $\forall V_i \in \Upsilon, V_Q \cap V_i \not= \phi$; 
		\emph{/*each entity or class vertex set $V_i$ has exactly one vertex in  $Q$*/}
		\item $|E_Q|=m$, and $\forall E_j \in \Gamma, E_Q \cap E_j \not= \phi$. 
		\emph{/*each predicate edge set $E_j$ has exactly one edge in $Q$*/}
	\end{enumerate}
	Each edge $e(\langle v_1, v_2\rangle,p) \in E_Q$ connects a pair of vertices $\langle v_1, v_2\rangle \in V_Q$ by a predicate $p$.
	The assembly cost of $Q$ is
	\begin{equation} \label{equ:cost}
	cost(Q)=\sum_{e(\langle v_1, v_2\rangle,p) \in E_Q}{w(\langle v_1, v_2\rangle,p)}
	\end{equation}
	where  $w(\langle v_1, v_2\rangle,p)$ denotes the triple assembly cost.

	The \emph{query graph assembly} (QGA for short) problem is to construct a valid graph $Q$ with the minimum assembly cost. 
\end{definition} 


There are two aspects that should be explained for QGA.

\textbf{\emph{1. Constraints:}}
The two constraints in Definition \ref{def:querygraphassembling} mean that each term $t^{v}_i$ ($1\leq i \leq n$) only corresponds to a single entity/class vertex in $Q$. For example, although ``USA'' may match two candidate entities dbo:USA$\_$Today and dbo:United$\_$States, in the final query graph $Q$, ``USA'' only matches a single entity (dbo:United$\_$States). It is analogue for the relation term $t_j^{e}$ ($1\leq j \leq m$). 

\textbf{\emph{2. Disengaged Edges:}}
A \emph{predicate edge} $e(\langle \cdot,\cdot \rangle,p)$ (in $E_j$) does not have two fixed endpoints but its edge label is fixed to predicate $p$. Thus, a predicate edge can be also called a \emph{disengaged} edge. The triple assembly cost $w(\langle v_1, v_2\rangle,p)$ measures the goodness of assembling $\langle v_1, v_2\rangle$ and $p$ into an edge in $Q$. Then the goal of the QGA problem is to determine the endpoints of $e(\langle \cdot,\cdot \rangle,p)$ to minimize the overall $cost(Q)$.

After finding the optimal $Q$ with minimum $cost(Q)$, we can translate it to SPARQL statements naturally, as illustrated in Figure \ref{fig:overview}.

\subsection{Graph Embedding Cost Model}
Note that the triple assembly cost $w(\langle v_1, v_2\rangle,p)$ can be any positive cost function, which does not affect the hardness of QGA. In other words, the QGA problem is a general computing framework to interpret the input keywords as SPARQL, which does not depend on any specific triple assembly cost function.

\begin{figure} [b]
	\centering
	\scalebox{0.55} [0.50]
	{
		\resizebox{\linewidth}{!}
		{
			\includegraphics[scale=1.0]{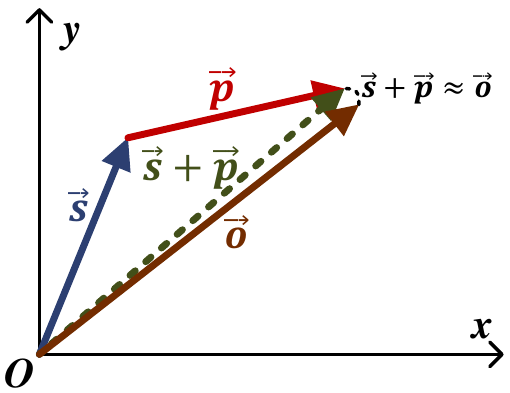}
		}
	}
	\caption{Visualizing the Intuition of Graph Embedding.}
	\label{fig:transe_visualizing}
	\vspace{-0.2in}
\end{figure}

The only thing affected by the selection of triple assembly cost function is the system's accuracy. A good cost function can guide to assemble correct query graph $Q$ that implies users' query intention. The process of assembling $\langle v_1, v_2\rangle$ and $p$ into a triple is analogue to ``link prediction'' problem in the RDF knowledge graph \cite{miller2009nonparametric}. Given two entity/class vertices $v_1$ and $v_2$, the link prediction is to ``predict'' the predicate $p$ between $v_1$ and $v_2$, and $w(\langle v_1,v_2\rangle, p)$ is a \emph{measure} of the prediction. Recent research show that the graph embedding technique is superior to other traditional approaches, such as \cite{miller2009nonparametric,nickel2011three,jenatton2012latent}. In the graph embedding model, all subjects (s), objects (o) and predicates (p) are encoded as multi-dimensional vectors $\overrightarrow{s}$, $\overrightarrow{p}$ and $\overrightarrow{o}$ such that $\overrightarrow s  + \overrightarrow p  \approx \overrightarrow o $ if $\langle s,p,o \rangle \in G$ (i.e., $\langle s,p,o\rangle$ is a triple in RDF graph); while $\overrightarrow s  + \overrightarrow p$ should be far away from $\overrightarrow o$ otherwise. Figure \ref{fig:transe_visualizing} visualizes the intuition. From the intuition, the structural information among entities, classes and relations in RDF graph is embedded into vectors. Therefore, we define the triple assembly cost based on graph embedding vectors as follows.

\begin{definition}\textbf{ (Triple Assembly Cost) }. \label{def:tripleassemblycost}
	Given two entity/class vertices $v_1$ and $v_2$ and a predicate edge $p$, the \emph{cost} of assembly triple $(v_1,p,v_2)$ is denoted as follows:
	\begin{equation}\label{equ:tripleassembly}
	w(\langle v_1, v_2\rangle, p) = MIN(| \overrightarrow{v_1} + \overrightarrow p  - \overrightarrow {v_2 } |, | \overrightarrow{v_2} + \overrightarrow p  - \overrightarrow {v_1 } |)
	\end{equation}
	where $\overrightarrow{v_1}$, $\overrightarrow {v_2}$ and $\overrightarrow {p}$ are the encoded multi-dimensional vectors  of $v_1$, $v_2$ and $p$, respectively. 
\end{definition}  

\begin{figure} [h]
	\centering
	\scalebox{1.0}
	{
		\resizebox{\linewidth}{!}
		{
			\includegraphics[scale=1.0]{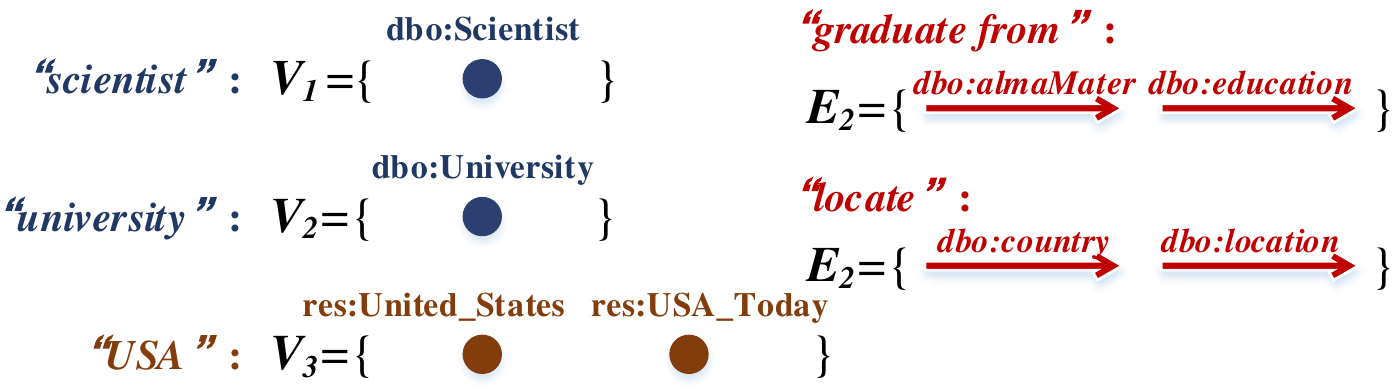}
		}
	}
	\caption{Elementary Query Graph Building Blocks.}
	\label{fig:graph_elements_exp2}
	\vspace{-0.3in}
\end{figure}

\begin{figure} [h]
	\newcommand{\mywidth}{0.23\textwidth}
	\centering
	\begin{subfigure}[t]{\mywidth}
		\centering
		\resizebox{\linewidth}{!}
		{
			\includegraphics{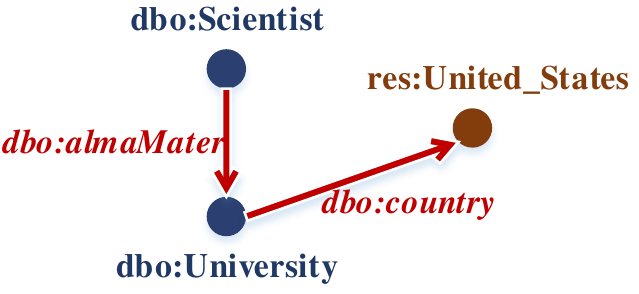}
			
		}
		\caption{$cost(Q_1)=1.76$}
		\label{fig:assembly_query_graph_q1}
	\end{subfigure}
	\begin{subfigure}[t]{\mywidth}
		\centering
		\resizebox{1.0\linewidth}{!}
		{
			\includegraphics{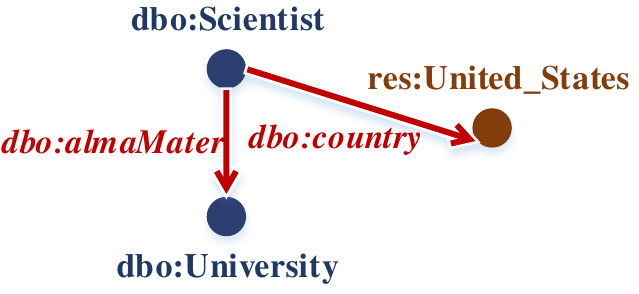}
		}
		\caption[font=\small]{$cost(Q_2)=2.46$}
		\label{fig:assembly_query_graph_q2}
	\end{subfigure}
	\caption{Possible Assembly Query Graphs.}
	\label{fig:assembly_query_graph}
	\vspace{-0.1in}
\end{figure}

\begin{example} In our example, there are three entity/class terms ``scientist'', ``university'', ``USA'' and two relation terms ``graduate from''  and ``locate''. Their corresponding entity/class vertices and predicate edges are shown in Figure \ref{fig:graph_elements_exp2}. There are two different assembly query graph $Q_1$ and $Q_2$ in Figure \ref{fig:assembly_query_graph}, among which $cost(Q_1)<cost(Q_2)$. Thus, the QGA problem result is $Q_1$ (Figure \ref{fig:assembly_query_graph_q1}).
\end{example}
\section{QGA: Hardness and Algorithm}
\label{sec:querygraphassembling}
Unfortunately, QGA is proved to be NP-complete in Section \ref{sec:hardness}. To solve that, we transform QGA into a constrained bipartite graph matching problem and design a practical efficient algorithm to find the optimal $Q$.  

\subsection{Hardness Analysis}\label{sec:hardness}

\begin{theorem}
	The query graph assembly problem is NP-complete.
\end{theorem}

\vspace{-0.1in}
\begin{proof}
	The decision version of QGA is defined as follows: 
	Given $n$ vertex sets $\Upsilon=\{V_1, ..., V_n\}$ and $m$ edge sets $\Gamma=\{E_1, ..., E_m\}$ and a threshold $\theta$, QGA is to decide if there exists an assembly query graph $Q$, where it satisfies the two constraints in Definition \ref{def:querygraphassembling} and the total assembly cost $cost(Q) \leq \theta$. Obviously, an instance of QGA can be verified in polynomial time. Thus, QGA belongs to NP class. 

	We construct a polynomial time reduction from 3-SAT (a classical NP-complete problem) to QGA. More specifically, given any instance $I$ of 3-SAT, we can generate an instance $I^{\prime}$ of QGA within polynomial time, where the decision value (TRUE/FALSE) of $I$ is \emph{equivalent} to $I^{\prime}$.
	
	\textbf{\emph{Any instance $I$ of 3-SAT:}}
	Without loss of generality, we define an instance $I$ of 3-SAT as follows: Given a set of $p$ boolean variables $U=\{u_1,u_2,...,u_p\}$ and a set of $q$ 3-clauses $C=\{c_1,c_2,...,c_q\}$ on $U$, The problem is to decide whether there exists a truth assignment for each variable in $U$ that satisfies all clauses in $C$.
	
	\textbf{\emph{Corresponding instance $I^{\prime}$ of QGA:}}
	Given a variable set $U$ and a 3-clause set $C$ (of instance $I$), we build a group of vertex sets $\Upsilon$ and a group of edge sets $\Gamma$ for the instance $I^{\prime}$.  $\Upsilon$ consists of the two parts $\Upsilon_{U}$ and $\Upsilon_{C}$, which are defined as follows:
	
	\begin{enumerate}
	\item For each variable $u_i \in U$, we introduce a vertex set $\{u_i,\overline{u_i}\}$ into $\Upsilon_{U}$. We call $u_i$ and $\overline{u_i}$ as \emph{variable vertices} of $I^{\prime}$. 
	\item For each 3-clause $c_j \in C$, we introduce a vertex set having a single vertex $\{c_j\}$ into $\Upsilon_{C}$. We call $c_j$ as a \emph{clause vertex} of $I^{\prime}$.  
	\end{enumerate}
	where $\Upsilon=\Upsilon_{U} \cup \Upsilon_{C}$.

	We also introduce $q$ disengaged edges into $\Gamma$. Each edge is a singleton set $\{e_j\}$. These $q$ edges can be used to connect any two vertices in $\Upsilon$. We set edge weight $w(e_j)=0$ if and only if $e_j$ connects the clause vertex $c_j$ and a member variable vertex $u_i$ (or $\overline{u_i}$) in 3-clause $c_j$. Otherwise, the edge weights are 1.
	
	The corresponding instance $I^{\prime}$ is defined as follows: Given two groups $\Upsilon$ and $\Gamma$ (explained above), and a threshold $\theta=0$, the problem is to decide whether there exists a graph $Q$, satisfying the two constraints in Definition \ref{def:querygraphassembling}, and $cost(Q)\leq 0$. Since edge weights are no less than 0, hence our goal is to construct $Q$ with $cost(Q)=0$.
	
	\textbf{\emph{Equivalence:}}
	Next, we will show the equivalence between the instance $I$ (of 3-SAT) and the instance $I^{\prime}$ (of QGA), i.e., $I \Leftrightarrow I^{\prime}$.

	Assume that the answer to $I$ is TRUE, which means that we have a truth assignment for each variable $u_i$ so that all 3-clauses $c_j$ are satisfied. According to the truth assignment in $I$, we can construct a graph $Q$ of $I^{\prime}$ as follows: for each clause vertex $c_j$, we connect $c_j$ with variable vertex $u_i$ (or $\overline{u_i}$) if $u_i=1$ (or $\overline{u_i}=1$, i.e, $u_i=0$) and $u_i$ is included in the 3-clause $c_j$. There may exist multiple variable vertices $u_i$  (or $\overline{u_i}$) satisfying the above condition. We connect $c_j$ with arbitrary one of them to form edge $(c_j,u_i)$ (or $(c_j,\overline{u_i})$).
	These edge weights are 0. 
	Thus, we construct a graph $Q$ satisfying the two constraints in Definition \ref{def:querygraphassembling} and $cost(Q)=0$.	It means the answer to $I^{\prime}$ is TRUE, i.e., $I \Rightarrow I^{\prime}$.
	
	Assume the answer to $I^{\prime}$ is TRUE. It means that we can construct a graph $Q$ having the following characters:
	\begin{enumerate}
		\item For $i=1,...,p$, one of variable vertex $\{u_i, \overline{u_i}\}$ is selected; and for $j=1,...,q$, clause vertex $c_j$ is selected; \emph{/*one vertex of each vertex set in $\Upsilon$ is selected*/}
		\item For $j=1,...,q$, each edge $e_j$ is selected; and $e_j$ connects the clause vertex $c_j$ and a variable vertex $u_i$(or $\overline{u_i}$) that corresponds to one of the three member variables of 3-clause $c_j$.  \emph{/*one edge of each edge set in $\Gamma$ is selected, and $cost(Q)=0$*/}
	\end{enumerate} 
	If a variable vertex $u_i$ (or $\overline{u_i}$) is selected, we set $u_i=1$ (or $\overline{u_i}=1$, i.e., $u_i=0$).  
	Since each clause vertex $c_j$ is connected to one selected variable vertex $u_i$ (or $\overline{u_i}$) that is in the 3-clause $c_j$, the corresponding variable $u_i=1$ (or $\overline{u_i}=1$). It means that 3-clause $c_j=1$ as $u_i$ (or $\overline{u_i}$) is included in $c_j$.  Hence, the answer to $I$ is also TRUE, i.e., $I \Leftarrow I^{\prime}$.
	
	In summary, we have reduced 3-SAT to QGA, where the former is a NP-complete problem. Therefore, we have proved that QGA is NP-complete. 

\end{proof}


\subsection{Assembly Bipartite Graph Model} \label{sec:general_case}
Since QGA is NP-complete, we transform it to an equivalent bipartite graph model with some constraints. Based on the bipartite graph model, we can design a best-first search algorithm with powerful pruning strategies.

Let us recall the definition of QGA (Definition \ref{def:querygraphassembling}), each of the $n$ entity/class terms is matched to a candidate vertex set $V_i$, and each of the $m$ relation terms is matched to a candidate edge set $E_j$. For example, ``USA'' is matched to $\{$res:USA$\_$Today, res:United$\_$States$\}$, and ``graduate from'' is matched to $\{$dbo:almaMater, dbo:education$\}$. The multiple choices in a candidate vertex/edge set indicate the ambiguity of keywords. If we adopt a pipeline style mechanism to address the keyword disambiguation and the query graph generation separately, we need to select exactly one element from each candidate vertex/edge set in the first phase. In our example, res:United$\_$States is the correct interpretation of ``USA'', and dbo:almaMater should be selected for ``graduate from''. If we simply adopt the string-based matching score \cite{li2011faerie} for keyword disambiguation, the matching score of res:USA$\_$Today may be higher than res:United$\_$States. In this case, if we only select the one with highest matching score, the correct answer will be missed due to the entity linking error. However, a robust solution should be error-tolerant with the ability to construct a correct query graph that is of interest to users even in the presence of noises and errors in the first phase. In our QGA solution, we allow the ambiguity of keywords (i.e., allowing one term matching several candidates) in the first phase, and push down the disambiguation to the query graph assembly step. For example, although the matching score of ``USA'' to res:USA$\_$Today is higher than that to res:United$\_$States, the former's assembly cost is much larger than the latter. Thus, we can still obtain the correct query graph $Q$. We propose the \emph{assembly bipartite graph matching} model to handle the ambiguity of keywords and the ambiguity of query graph structures uniformly. 

\begin{definition}\textbf{ (Assembly Bipartite Graph) }. \label{def:assemblygraph}
	Each entity/class term $t_i^{v}$ corresponds to a set $V_i$ of vertices ($1\leq i \leq n$) and each relation term $t_j^{e}$ corresponds to a set $E_j$ of predicate edges ($1\leq j \leq m$).  
	
	An assembly bipartite graph $\mathbb{B}(V_{L},V_{R},E_{\mathbb{B}})$ is defined as follows:
	
	\begin{enumerate}
		\item Vertex pair set $V_{i_1} \times V_{i_2}=\{(v_{i_1},v_{i_2}) | 1\leq i_1 < i_2 \leq n \wedge v_{i_1} \in V_{i_1}  \wedge v_{i_2} \in V_{i_2}\}$.
		\item $V_{L}=\bigcup\nolimits_{1 \le i_1  < i_2  \le n } {(V_{i_1} \times V_{i_2})} $.
		\item  $V_{R}=\bigcup\nolimits_{1 \le j \le m} {E_j }  $.
		\item there is a crossing edge $e$ between any node $(v_{i_1}, v_{i_2})$ in $V_{L}$ and any node $p_j$ in $V_{R}$ ($1 \le i_1  < i_2  \le n$, $1 \le j \le m$), which is denoted as $e(\langle v_{i_1},v_{i_2}\rangle, p_j)$. Edge weight $w(e)=w(\langle v_{i_1},v_{i_2}\rangle, p_j)$, where $w(\langle v_{i_1},v_{i_2}\rangle, p_j)$ denotes the triple assembly cost.  
	\end{enumerate} 
\end{definition}

\begin{figure} [t]
	\begin{center}
		\vspace{-0.15in}
		\includegraphics[scale=0.55]{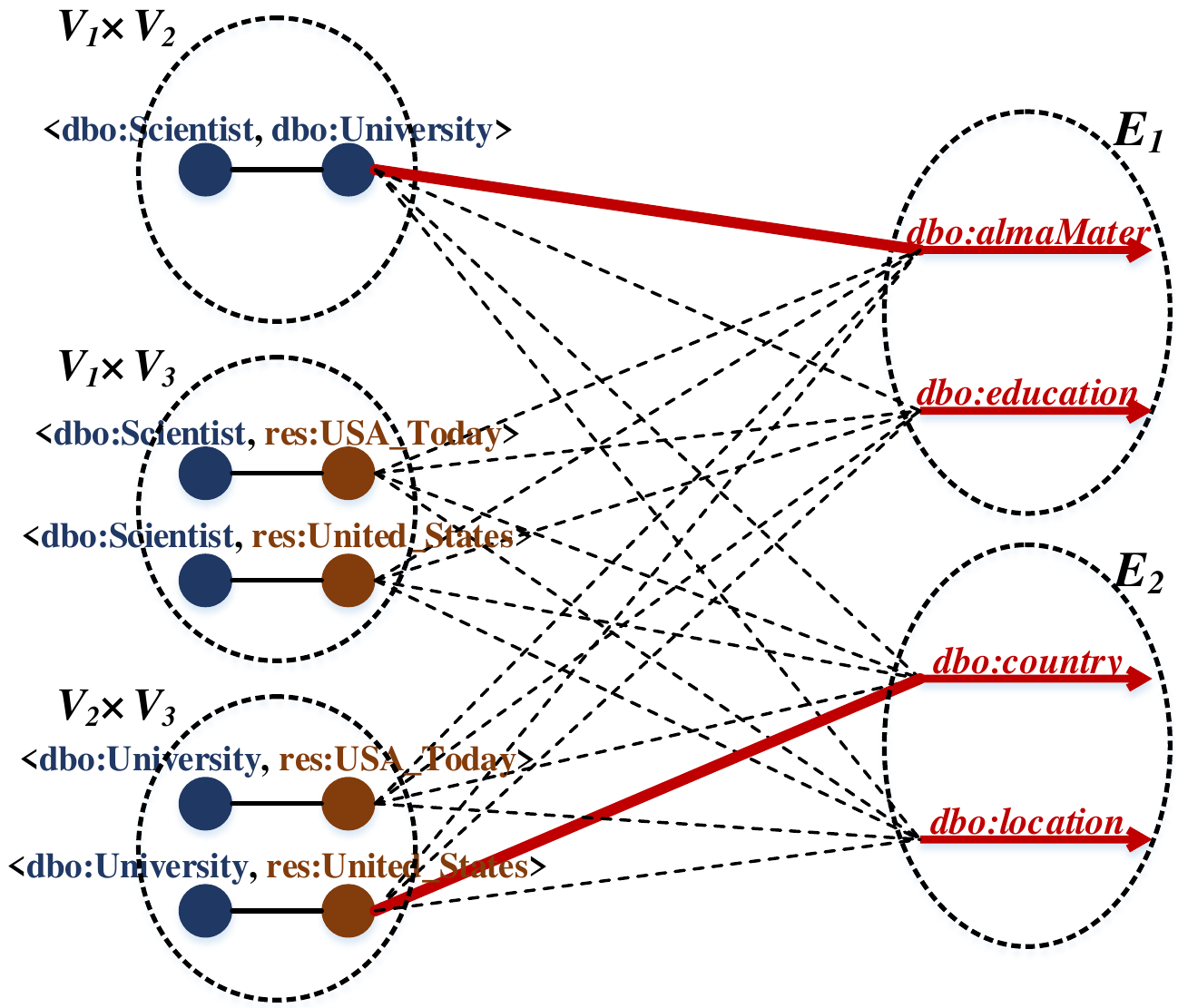}
		\caption{An Example of Assembly Bipartite Graph.}
		
		\label{fig:asm_bigraph}
		\vspace{-0.15in}
	\end{center}
\end{figure}

\begin{example} 
Let us recall the running example. Each term's candidate matchings are given in Figure \ref{fig:graph_elements_exp2}, i.e., $V_1=\{$dbo:Scientist$\}$, $V_2=\{$dbo:University$\}$, and $V_3=\{$res:United\_Today, res:United\_States$\}$. Thus, there are three vertex pair sets $V_1 \times V_2$, $V_2 \times V_3$ and $V_1 \times V_3$. In Figure \ref{fig:asm_bigraph}, each vertex pair set is highlighted by a dash circle. There are two predicate edge sets $E_1=\{$dbo:almaMater, dbo:education$\}$ and $E_2=\{$dbo:country, dbo:location$\}$, which are also illustrated in Figure \ref{fig:asm_bigraph} using dash circles. 
\end{example}

It is worth noting that there are some \emph{conflict} relations among the crossing edges in $\mathbb{B}$.  For example, crossing edge $e(\langle$dbo:Scientist, res:USA\_Today$\rangle$, dbo:country) conflicts with $e^{\prime}(\langle$dbo:University, res:United\_States$\rangle$, dbo:almaMater) (in Figure \ref{fig:asm_bigraph}), since the semantic term ``USA'' corresponds to two different entity vertices res:USA\_\\Today and res:United\_States in $e$ and $e^{\prime}$, respectively. In this case, the constraints of QGA (in Definition \ref{def:querygraphassembling}) will be violated if $e$ and $e^{\prime}$ occur in the same matching.
Considering the above example, we formulate the \emph{conflict relation} among crossing edges in $\mathbb{B}$. 

\begin{definition} \textbf{ (Conflict Relation) }.\label{def:conflict}	
	For any two crossing edges $e(\langle v_{i_1},v_{i_2}\rangle, p_j)$ and $e^{\prime}(\langle v_{i_1}^{\prime},v_{i_2}^{\prime}\rangle, p_j^{\prime})$ in an assembly bipartite graph $\mathbb{B}$, we say that $e$ is \emph{conflict} with $e^{\prime}$ if at least one of the following conditions holds:
	
	\begin{enumerate}
		\item $v_{i_1}$ and $v_{i_1}^{\prime}$ (or $v_{i_2}$ and $v_{i_2}^{\prime}$) come from the same vertex set $V_{i_1}$ (or $V_{i_2}$) and  $v_{i_1} \neq v_{i_1}^{\prime}$ (or $v_{i_2} \neq v_{i_2}^{\prime}$);		
		\ /*Two different vertices come from the same vertex set; an entity/class term $t_i^{v}$ cannot map to multiple vertices*/
		\item $p_j$ and $p_j^{\prime}$ come from the same edge set $E_j$ and $p_j \neq p_j^{\prime}$.		
		\ /*Two different predicates come from the same edge set; an relation term $t_j^{e}$ cannot map to multiple predicate edges*/
		\item $(v_{i_1}=v_{i_1}^{\prime} \wedge v_{i_2}=v_{i_2}^{\prime}) \vee (p_j = p_j^{\prime})$ 		
		\ /*$e$ and $e^{\prime}$ share a common endpoint in $V_{L}$ or $V_{R}$; a vertex pair cannot be assigned two different predicate edges and a predicate edge cannot connect two different vertex pairs */
	\end{enumerate}

\end{definition}

In order to be consistent with the constraints of QGA, we also redefine \emph{matching} as follows.
\begin{definition}\textbf{ (Matching) }\label{def:matchingnew}. Given an assembly bipartite graph $\mathbb{B}(V_{L},V_{R},E_{\mathbb{B}})$, a \emph{matching} of $\mathbb{B}$ is a subset $M$ ($M \subseteq E_{\mathbb{B}}$) of its crossing edges, no two of which are \emph{conflict} with each other.
\end{definition}

\begin{definition}\textbf{ (Matching Cost) }. The \emph{matching cost} of $M$ is $w(M) = \sum\nolimits_{e \in M} {w(e)}$, where $w(e)$ is defined in Definition \ref{def:assemblygraph}(4). 
\end{definition}
It is easy to know that solving QGA problem is equivalent to finding a size-$m$ \emph{matching} over $\mathbb{B}$ with the minimum \emph{matching cost}. A matching edge in $\mathbb{B}$ stands for an assembled edge in $Q$.

\subsection{Condensed Bipartite Graph}

\begin{figure} [t]
	\begin{center}
		\includegraphics[scale=0.65]{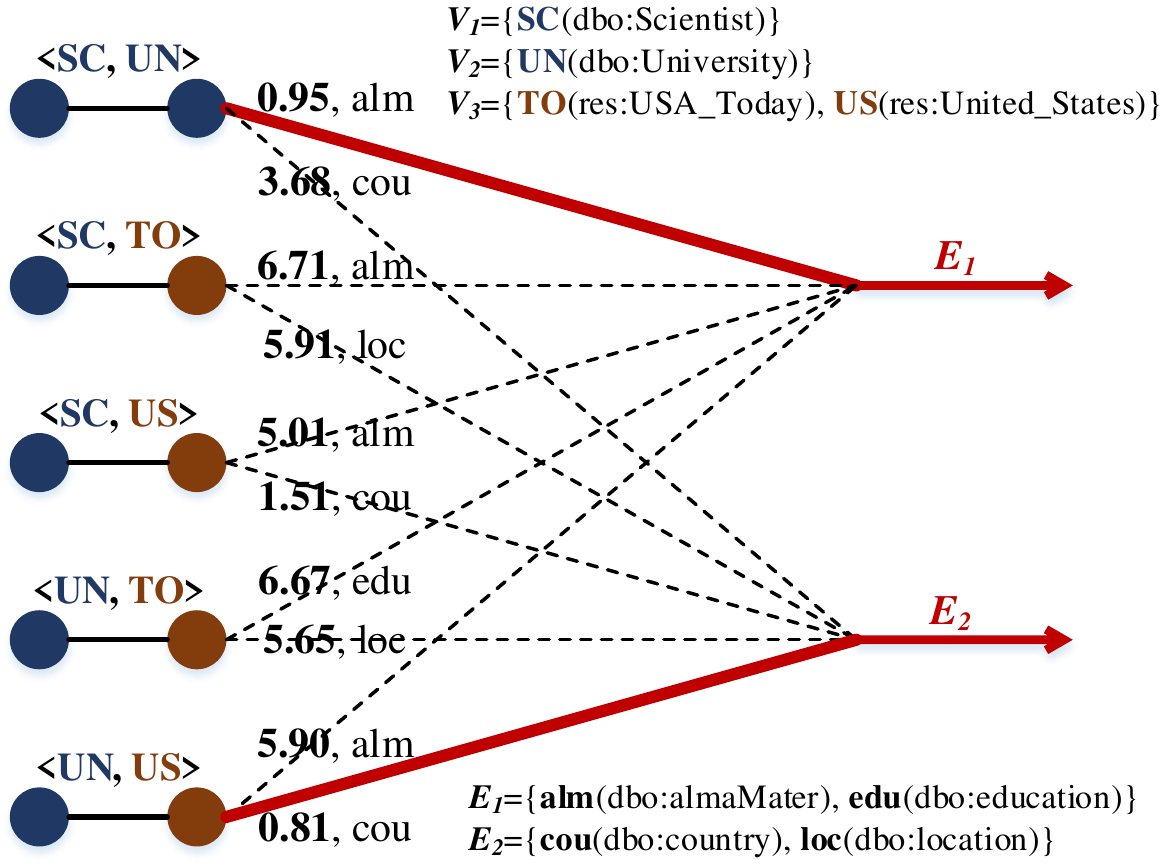}
		\caption{An Example of Condensed Bipartite Graph.}
		\label{fig:con_bigraph}
		   \vspace{-0.2in}
	\end{center}
\end{figure}

According to the condition (2) and (3) in Definition \ref{def:conflict}, each predicate edge set $E_j$ has only one $p_j$ occurring in the \emph{matching} $M$. Thus, we can condense each predicate edge set into one node $E_j$, leading to a \emph{condensed bipartite graph} $\mathbb{B}^{*}(V_{L}^*,V_{R}^*,E_{\mathbb{B}}^*)$, as shown in Figure \ref{fig:con_bigraph}. There is a crossing edge between any node $(v_{i_1},v_{i_2})$ in $V_{L}^*$ and any node $E_j$ in $V_{R}^*$. The edge weight is defined as $w(e) = w(\langle v_{i_1} ,v_{i_2}\rangle ,E_j ) = MIN_{p_j  \in E_j } \{ w(\langle v_{i_1} ,v_{i_2}\rangle,p_j)\}$, where $w(\langle v_{i_1 } ,v_{i_2}\rangle,p_j)$ is defined in Definition \ref{def:assemblygraph}(4). In order to find the size-$m$ matching with the minimum cost in $\mathbb{B}^{*}$, we propose QGA Algorithm (i.e., Algorithm \ref{alg:bfmatch}).

\subsection{QGA Algorithm}

A search state is denoted as $\{M,Z, cost(M), LB(M)\}$, where $M$ records the current partial matching, i.e., a set of currently selected matching edges, $Z$ records a set of unmatched edges that are not \emph{conflict} with edges in $M$. Obviously, each edge $e \in Z$ can be appended to $M$ to enlarge the current matching. Initially, $M=\phi$, and $Z$ records all crossing edges in the condensed bipartite graph $\mathbb{B}^{*}$ (Line 3 in Algorithm \ref{alg:bfmatch}). $cost(M)$ denotes the current partial matching cost, i.e., $cost(M)= \sum\nolimits_{e \in M} {w(e)}$. $LB(M)$ denotes the lower bound of the current partial matching $M$. We will discuss how to compute the lower bound $LB(M)$ later. All search states are stored in a priority queue $H$ in the non-decreasing order of the lower bound $LB(M)$ (Line 2). Furthermore, we maintain a threshold $\theta$ to be the current minimum matching cost. Initially, $\theta = \infty$. In each iteration, we pop a head state $\{M,Z, cost(M), LB(M)\}$.
We enumerate all unmatched edges $e \in Z$ in non-decreasing order of $w(e)$ to generate subsequent search states. Each $e \in Z$ is moved from $Z$ to $M$ to obtain a new matching $M^{\prime}$ (Line 9). We remove $e$ and all edges in $Z$ that are conflict with $e$ to obtain $Z^{\prime}$ (Line 10). We also update $cost(M^{\prime})$ and $LB(M^{\prime})$ (Lines 11-12). Then we check whether $M^{\prime}$ is a size-$m$ matching over $\mathbb{B}^{*}$(i.e. end state). If so, we update the threshold $\theta$ if $cost(M^{\prime}) < \theta$ and record $M^{\prime}$ as the current optimal matching $M_{opt}$  (Lines 13-16). Otherwise, we push the new state $\{M^{\prime},Z^{\prime}, cost(M^{\prime}), LB(M^{\prime})\}$ into $H$ (Line 18). The algorithm keeps iterating until that the threshold $\theta$ is less than the lower bound of the current head state in $H$ (Line 6) or $H$ is empty. 
\begin{algorithm} 
	\caption{QGA Algorithm} \label{alg:bfmatch}
	\KwIn{Condensed bipartite graph $B^*(V_{L}^*, V_{R}^*, E_B^*)$,
		and conflict relations among $E_B^*$.}
	\KwOut{The optimal assembly query graph $Q$.}
	$M_{opt} := \phi$, $\theta := \infty$ \;
	$H := \phi$ ;\tcp{min-heap, sort by lower bound}
	$H \leftarrow \{M:=\emptyset, Z := E_B^*, cost(M) := 0, LB(M) := 0\}$ \;
	\While{$H$ is not empty}
	{
		$\{M, Z, cost(M), LB(M)\} \leftarrow H.pop$ \;
		\If{$LB(M) >= \theta$}{\textbf{break} \;}
		
		\For{each crossing edge $e \in Z$}
		{
			$M^{\prime} := M \cup \{e\}$; $Z := Z \setminus \{e\}$ \;
			$Z^{\prime} = Z \setminus \{e^\prime | e^\prime \in Z \wedge e^\prime\ conflict\ with\ e\}$ \;
			$cost(M^{\prime}) := cost(M) + w(e)$ \;
			Compute $LB(M^{\prime})$\;
			
			\If{$|M^{\prime}| = m$}
			{
				\If{$cost(M^{\prime}) < \theta$}
				{
					$\theta := cost(M^{\prime})$ \;
					$M_{opt} := M'$ \;
				}       
			}
			\Else
			{
				$H \leftarrow \{M', Z', cost(M^{\prime}), LB(M^{\prime})\}$ \;
			}       
		}
	}
	Build query graph $Q$ according to $M_{opt}$ \;
	\Return $Q$
\end{algorithm}

\vspace{-0.1in}
\subsection{Computing Lower Bound} \label{sec:lower_bound}
Considering a search state $\{M, Z, cost(M), LB(M)\}$, we discuss how to compute $LB(M)$. Let $|M|$ denote the number of edges in the current matching. Since a maximum matching in $B^{*}$ should contain $m$ matching edges (i.e., covering all $E_j$), we need to select the other $(m-|M|)$ no-conflict edges from $Z$, to form a size-$m$ matching. We denote  $\mathcal{M}(Z)$ as the minimum weighted size-$(m-|M|)$ matching of $Z$. Thus, the best result that can be reached from $M$ is $M \cup \mathcal{M}(Z)$. To ensure the correctness of Algorithm \ref{alg:bfmatch}, $LB(M) \le cost(M \cup \mathcal{M}(Z))$ must always be satisfied. A good lower bound should have the following two characters: (1)  $LB(M)$ is as close to $cost(M \cup \mathcal{M}(Z))$ as possible; (2) the computation cost of $LB(M)$ is small. From this intuition, we propose three different lower bounds as follows.

\Paragraph{Naive-LB:}
The naive method to compute $LB(M)$ is to select the top-$(m-|M|)$ unmatched edges $\{e_1,...,e_{m-|M|}\}$ in $Z$ with the minimum weights and compute $Navie$-$LB(M) = cost(M) + \sum\nolimits_{i = 1}^{i = m  - |M|} {w(e_i)}$. In the implementation of Algorithm \ref{alg:bfmatch}, $Z$ is stored by a linked list, and always kept in the non-decreasing order of $w(e)$. Therefore, the complexity of computing Naive-LB is $O(m-|M|)$. Since the top-$(m-|M|)$ unmatched edges in $Z$ may be conflict with each other, Naive-LB is not tight, and has the weakest pruning power comparing with the following two lower bounds.

\Paragraph{KM-LB:}
We ignore the conflict relations among unmatched edges in $Z$, and find the minimum weighted size-$(m-|M|)$ matching $\mathcal{M}_{KM}(Z)$ by KM Algorithm \cite{kuhn1955hungarian} in $O(|Z|^3)$ time. Then we compute $KM$-$LB(M)= cost(M) + cost(\mathcal{M}_{KM}(Z))$. KM-LB is much tighter than Naive-LB, but the computation cost is expensive.

\Paragraph{Greedy-LB:}
Inspired by the matching-based KM-LB, we propose a greedy strategy (Algorithm \ref{alg:greedy_lb}) to find an approximate matching of $Z$ in $O(|Z|)$ time. It has been proved in \cite{preis1999linear} that the result return by Algorithm \ref{alg:greedy_lb} (denote as $\mathcal{M}_{greedy}(Z)$) is a $1/2$-approximation of the optimal matching $\mathcal{M}_{KM}(Z)$. i.e. $cost(\mathcal{M}_{KM}(Z)) \ge \frac{cost(\mathcal{M}_{greedy}(Z))}{2}$. Thus we compute $Greedy$-$LB(M)= cost(M) + \frac{cost(\mathcal{M}_{greedy}(Z))}{2}$. Greedy-LB is considered as the trade-off between Naive-LB and KM-LB, because of its medium tightness and computation cost. Experiment in Section \ref{sec:evaluate_qga_bounds} confirms that Greedy-LB gains the best performance among them.

\begin{algorithm} [t]
	\caption{Greedy-LB Algorithm} \label{alg:greedy_lb}
	\KwIn{Unmatched edge set $Z$.}
	\KwOut{An $1/2$-approximate matching $\mathcal{M}_{greedy}$.}
	$\mathcal{M}_{greedy} := \phi$ \;
	\For{each $e \in Z$ in non-decreasing order of $w(e)$}
	{
		\If{$e$ do not share common vertices with $\mathcal{M}_{greedy}$}
		{
			$\mathcal{M}_{greedy} := \mathcal{M}_{greedy} \cup \{e\}$
		}
		\Return $\mathcal{M}_{greedy}$ \;
	}
\end{algorithm}

\vspace{-0.1in}
\subsection{Time Complexity Analysis} 
In a condensed bipartite graph $\mathbb{B}^{*}(V_{L}^*,V_{R}^*,E_{\mathbb{B}}^*)$, $|V_L^*| \le k^2 \cdot \left( \begin{array}{l}  n \\  2 \\  \end{array} \right)$, because there are $\left( \begin{array}{l}  n \\  2 \\  \end{array} \right)$ vertex pair set, each contains at most $k^2$ vertex pairs, and $V_R^*$ consists of $m$ condensed predicate nodes, i.e., $|V_R^*|=m$. Therefore, the total number of crossing edges is $|E_{\mathbb{B}}^*|=|V_R^*|\times|V_L^*| \le mk^2\left( \begin{array}{l}  n \\  2 \\  \end{array} \right) \le mn^2k^2$. Since a maximum matching in $\mathbb{B}^{*}$ should contain $m$ matching edges, we can find at most $\left( \begin{array}{l}  mn^2k^2 \\  \ \ \ m \\  \end{array} \right) \le (mn^2k^2)^m$ different maximum matchings. Since $n+m \le l$, by replacing $n$ and $m$ by $l$, we get the overall search space $O(k^{2l} \cdot l^{3l})$.

\subsection{Implicit Relation Prediction}
As we know, keywords are more flexible than NL question sentences. In some cases, users may omit some relation terms. For example, users may input ``scientist graduate from university USA'', where the keyword ``locate'' is omitted. In this case, for humans, it is trivial to infer that the user means ``an university located in USA''. Let us recall our QGA approach. If we omit ``locate'' in the running example, there is only one relation term. So, the query graph $Q$ cannot be connected if we use only one predicate edge to connect three vertices. We can patch our solution to connect $Q$ as follows.

\begin{definition} \textbf{ (Relation Prediction Graph) }
	Suppose that $Q$ consists of $r$ connected components: $Q=\{\mathcal{P}_1, \mathcal{P}_2, ..., \mathcal{P}_r\}$. A relation prediction graph $P(V_P,E_P)$ is a complete graph and defined as follows:
	\begin{itemize}
		\item $V_P$ consists of $r$ vertices, where each vertex $v_P^i \in V_P$ corresponds to a connected component $\mathcal{P}_i$.
		\item $E_P$ consists of $\frac{r(r-1)}{2}$ labeled weighted edges. For any $(\mathcal{P}_i, \mathcal{P}_j)$, we find the minimum assembly cost triple $e(\langle v_i, v_j \rangle, p)$, where $v_i \in V(\mathcal{P}_i)$, $v_j \in V(\mathcal{P}_j)$, and $p \in \mathcal{U}$. Note that $V(\mathcal{P}_i)$ denotes all vertices in $P_i$, and $\mathcal{U}$ denotes all predicates in RDF graph $G$. Then we add a corresponding edge $e_P(v_P^i,v_P^j)$ into $E_P$, where the weight $w(e_P)=w(\langle v_i, v_j \rangle, p)$, and the label $l(e_P)=(\langle v_i, v_j \rangle, p)$.
	\end{itemize}
\end{definition}
Thus the relation prediction task is modeled as finding a minimum spanning tree $T$ on $P$, which can be solved in linear time. Through the label $(\langle v_i, v_j \rangle, p)$ of the tree edge in $T$, we can know that $v_i, v_j \in V_Q$ should be connected by the predicate edge $p$.

\section{Evaluation}\label{sec:exp}

We evaluate both effectiveness and efficiency of our approach, and compare with a typical data graph based approach DPBF \cite{ding2007finding}, and a summary graph based approach SUMG \cite{tran2009top}, which are mentioned in Section \ref{sec:related_work}.
Due to the lack of schema information in RDF graphs, we can not compare with schema graph based approaches. Our system participates the QALD-6 competition and we report the results of our system with regard to other NL-QA participants. Experimental studies are also conducted on Freebase. All implementations are in Java, and all experiments are performed on a Linux server with Intel Xeon E5-2640v3@2.6GHz, 128GB memory, and 4T disk.

\vspace{-0.1in}
\subsection{Dataset}

\textbf{DBpedia + QALD-6:}
QALD-6 \cite{unger20166th} collects 100 queries over DBpedia.
For each query, it provides both NL question sentences and keywords.
we only take each query's keywords as the input to our system, DPBF, and SUMG, while other QALD-6 participants can leverage both of NL sentences and keywords.


\noindent\textbf{Freebase + Free917:}
Free917 \cite{cai2013large} is an open QA benchmark which consists of NL question and answer pairs over Freebase. We select 80 questions of it and rewrite them into keyword queries artificially. 

Table \ref{tab:statistic_rdf_graph} lists the statistics of the two RDF datasets.
\begin{table} [h]
\begin{tabular}
{  r | c | c }
\hline
 & DBpedia & Freebase \\ 
\hline
entities & 6.7M & 153M \\
relations & 1.4K & 19K\\
classes & 0.6K  & 15K\\
total triples & 583M & 1.9B \\
\hline
\end{tabular}
\caption{Statistics of RDF Datasets.}
\label{tab:statistic_rdf_graph}
\vspace{-0.3in}
\end{table}

%
%
%

\noindent\textbf{TransE Vectors:}
%
We adopt an open-sourced TransE implementation
\footnote{https://github.com/thunlp/Fast-TransX} to train vectors.
We use a grid search strategy to find an optimal parameter configuration: learning rate $\lambda=10^{-5}$, number of iterations $n=3000$, vector dimension $d=200$. The training process costs 2.1h and 5.8h on DBpedia and Freebase, respectively.

\vspace{-0.1in}
\subsection{Evaluating Effectiveness}

\begin{table} [t]
	\scalebox{1.0}{
		\begin{tabular}
			{ l | r | c c c c}
			\hline
			& & \textbf{N} & \textbf{R} & \textbf{P} & \textbf{F-1} \\
			\hline
			\multirow{5}{*}{\begin{minipage}{0.5in} Keyword Search Systems\end{minipage}}
			& \textbf{\textit{QGA(+TransE)}} & 100 & 0.59 & 0.85 & \textbf{\textit{0.70}} \\ 
			& \textit{QGA(+Cooccur)}          & 100 & 0.49 & 0.76 & 0.59 \\
			& \textit{QGA(+TFIDF)}           & 100 & 0.37 & 0.64 & 0.47 \\			
			& DPBF                         & 100 & 0.46 & 0.21 & 0.29 \\
			& SUMG                         & 100 & 0.31 & 0.22 & 0.23 \\  
			\hline
			\multirow{6}{*}{\begin{minipage}{0.5in} NL-QA Systems\end{minipage}}
			& CANaLI                       & 100 & 0.89 & 0.89 & 0.89 \\ 
			& UTQA                         & 100 & 0.69 & 0.82 & 0.75 \\ 
			& \textbf{\textit{QGA(+TransE)}} & 100 & 0.59 & 0.85 & \textbf{\textit{0.70}} \\  
			& NBFramework                  & 63  & 0.54 & 0.55 & 0.54 \\ 
			& SemGraphQA                   & 100 & 0.25 & 0.70 & 0.37 \\ 
			& UIQA                         & 44  & 0.28 & 0.24 & 0.25 \\ 
			\hline
		\end{tabular}
	}
	\begin{tablenotes}
		\item \textbf{N}: Number of Processed Queries; \textbf{R}: Recall; \textbf{P}: Precision.
	\end{tablenotes}
	\caption{QALD-6 Evaluation Result}
	\label{tab:testres}
	\vspace{-0.3in}
\end{table}
\begin{table} [h]
	\begin{tabular}
		{r | c c c c}
		\hline
		& \textbf{N} & \textbf{R} & \textbf{P} & \textbf{F-1} \\
		\hline
		\textbf{\textit{QGA(+TransE)}} & 80 & 0.62 & 0.65 & \textbf{\textit{0.63}} \\
		\textit{QGA(+Cooccur)}          & 80 & 0.32 & 0.55 & 0.40 \\
		DPBF                         & 80 & 0.55 & 0.28 & 0.37 \\
		\textit{QGA(+TFIDF)}           & 80 & 0.26 & 0.52 & 0.35 \\
		SUMG                         & 80 & 0.24 & 0.19 & 0.21 \\
		\hline
	\end{tabular}
	\caption{Free917 Evaluation Result}
	\label{tab:freebaseres}
	\vspace{-0.4in}
\end{table}

\subsubsection{Overall Effectiveness of Our Approach}
For the keyword search systems, we compare our approach with DPBF and SUMG ,in Table \ref{tab:testres} and \ref{tab:freebaseres}. DPBF uses a dynamic programming strategy to find top-k GST over the data graph, and returns the tree nodes as answers. In DPBF, the answer set's size influences its accuracy significantly, because it varies widely on different queries.
To favor the comparison, we assume that DPBF knows the answer set's size in advance. For each query, if its standard answer set's size is $|R|$, we set DPBF to return $3 \times |R|$ candidate answers. Then we calculate different F-1 by resizing the answer set from $|R|$ to $3\times |R|$ and find the highest one as the measure value. Despite such a favorite measurement for DPBF, it still cannot achieve desirable accuracy.

SUMG explores the subgraph that connects keywords' mapping candidates with a minimum cost over a class based summary graph $G_S$ \cite{tran2009top}.
Because $G_S$ loses detailed information about relations between entities of the same class, SUMG performs even worse than DPBF in accuracy.


We also compare our approach with some NL-QA systems participating QALD-6. Table \ref{tab:testres} lists the top-6 systems that are ranked by F-1. Our system ranks at top-3 even that we only use keywords instead of the complete NL question sentences. Actually, the top-1 system CANaLI \cite{atzori2016answering} requires manual efforts for disambiguation. Our approach is better than the rank-2 system UTQA in the precision, but the recall of our approach is worse than that. As mentioned before, NL-QA system can leverage the syntactic structure of the question sentences, but it is unavailable for the keyword search system. Even so, our approach is still competitive with these NL-QA systems in effectiveness.

\vspace{-0.05in}
\subsubsection{Comparing with Different Assembly Metrics}
To show the superiority of graph embedding metrics, we also compare with traditional link prediction methods in Table \ref{tab:testres} and \ref{tab:freebaseres}. Given a triple $\langle s,p,o \rangle$, \emph{Cooccur} measures the co-occurrence frequency of $\langle s,p \rangle$, $\langle p,o \rangle$, and $\langle p,o \rangle$. \emph{TFIDF} adopts the tf-idf formula to measure the selectivity of $p$ on $\langle s, o \rangle$. As we can see, both of the two metrics are inferior to TransE on accuracy, because they can hardly capture the structural information of the RDF graph.

\subsubsection{Tuning Parameter $N$ and $k$}

\begin{figure} [h]
	\newcommand{\mywidth}{0.22\textwidth}
	\centering
	\begin{subfigure}[t]{\mywidth}
		\centering
		\scalebox{0.45}
		{
			\includegraphics{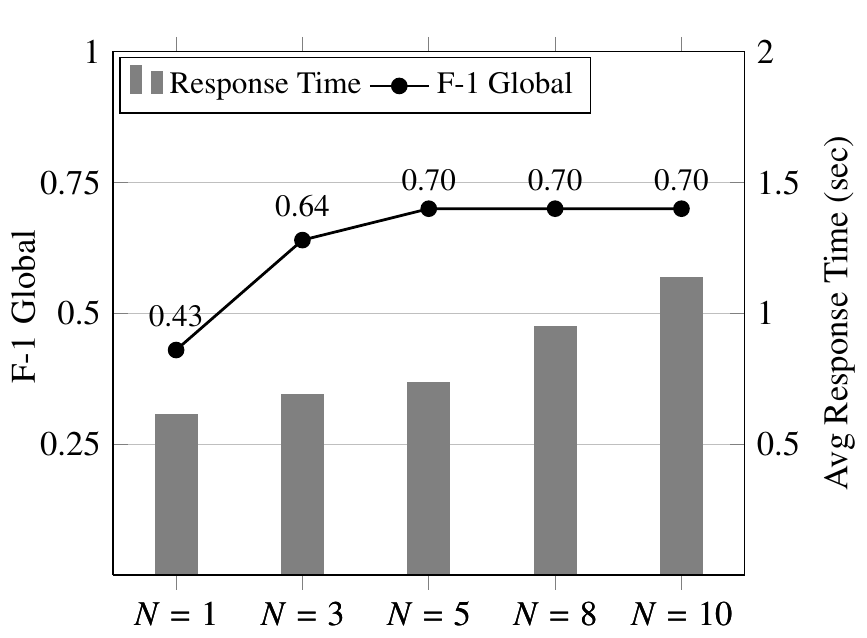}
			
		}
		\vspace{-0.2in}
		\caption{DBpedia}
		\label{fig:chooseoptn_dbpedia}
	\end{subfigure}
	\begin{subfigure}[t]{\mywidth}
		\centering
		\scalebox{0.45}
		{
			\includegraphics{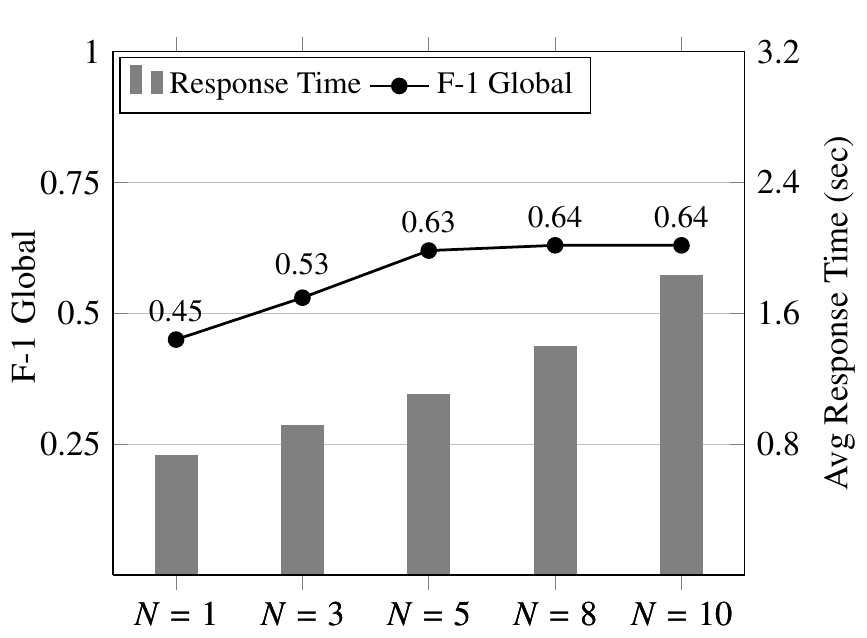}
		}
		\vspace{-0.2in}
		\caption{Freebase}
		\label{fig:chooseoptn_freebase}
	\end{subfigure}
	\vspace{-0.1in}
	\caption{Tuning Candidate AQ Size $N$}
	\vspace{-0.2in}
	\label{fig:chooseoptn}
\end{figure}

In Section \ref{sec:phase1}, we mentioned that a keyword token sequence $RQ$ will be interpreted as top-$N$ $AQ$.
Along with $N$ growing larger, Phase-II needs to process more $AQ$, leading to a trade-off between system's accuracy and response time. Thus we tune $N \in [1,10]$  in Figure \ref{fig:chooseoptn}, and find that the growth of F-1 is almost stagnant when $N$ is larger than $5$.

\begin{figure} [h]
	\newcommand{\mywidth}{0.22\textwidth}
	\centering
	\begin{subfigure}[t]{\mywidth}
		\centering
		\scalebox{0.45}
		{
			\includegraphics{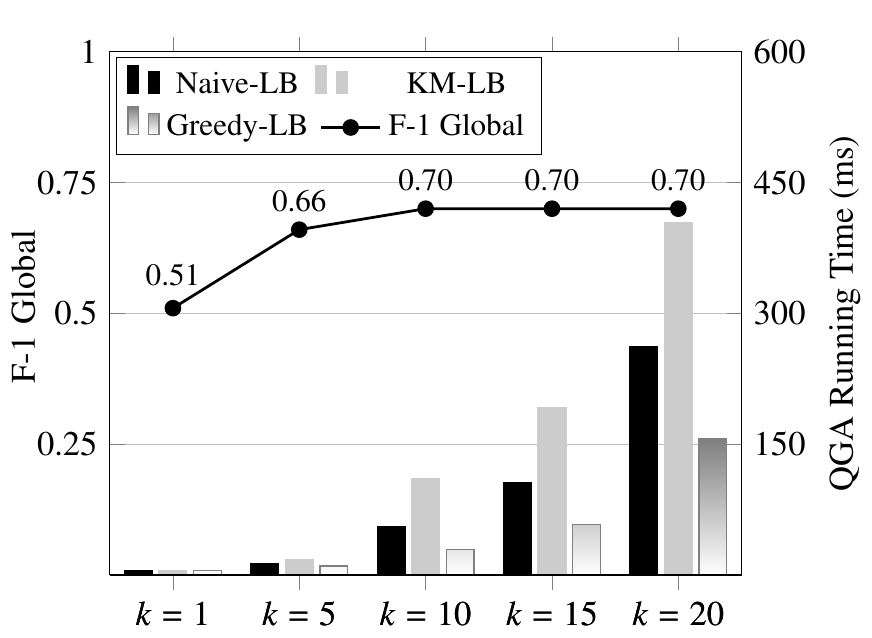}
			
		}
		\vspace{-0.2in}
		\caption{DBpedia}
		\label{fig:lb_f1_dbpedia}
	\end{subfigure}
	\begin{subfigure}[t]{\mywidth}
		\centering
		\scalebox{0.45}
		{
			\includegraphics{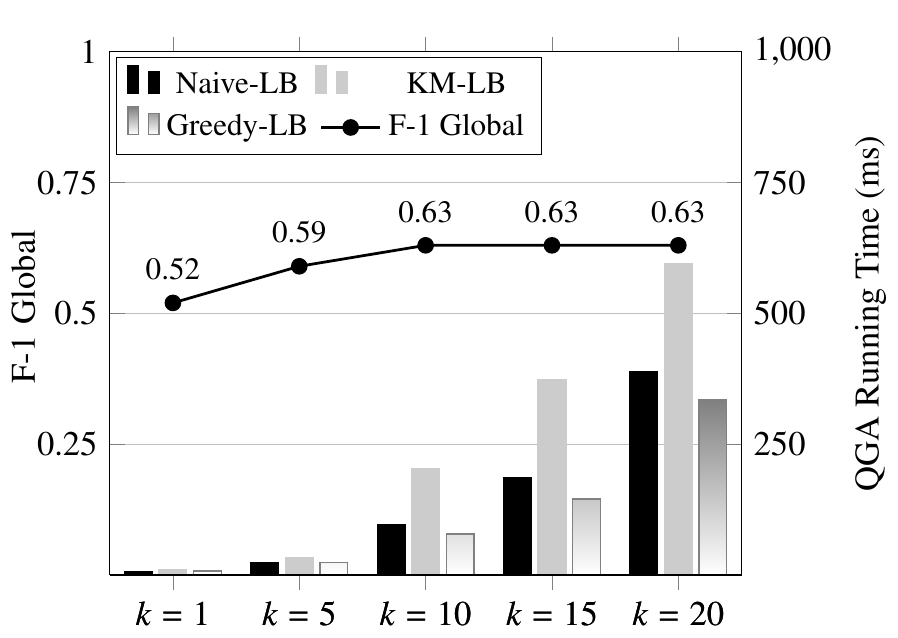}
		}
		\vspace{-0.2in}
		\caption{Freebase}
		
		\label{fig:lb_f1_freebase}
	\end{subfigure}
	\vspace{-0.1in}
	\caption{Tuning $k$ and Comparing Different Lower Bounds}
	\label{fig:lb_f1}
	\vspace{-0.1in}
\end{figure}

In QGA, we suppose that each term is matched to at most $k$ query graph elements. Similarly, threshold $k$ also leads to a trade-off between precision and performance. Thus we tune $k \in [1,20]$  in Figure \ref{fig:lb_f1}, and find that the growth trend of F-1 will stop when $k$ is larger than $10$, while the running time still raises. Therefore, we set $N=5$ and $k=10$ in the comparative experiments.

\vspace{-0.1in}

\subsection{Evaluating Efficiency} \label{sec:efficiency}

\subsubsection{Evaluating QGA Algorithm and Lower Bounds} \label{sec:evaluate_qga_bounds}
We evaluate the efficiency of QGA Algorithm, with three different lower bounds proposed in Section \ref{sec:lower_bound}. We count the average number of search states under the three lower bounds in Table \ref{tab:search_state_cnt}. The QGA running time with different lower bounds is illustrated in Figure \ref{fig:lb_f1}. We can see that KM-LB gives us the tightest lower bound, but it runs slowest, because the lower bound computation by KM Algorithm is expensive. Both of Naive-LB and Greedy-LB have linear time complexity, while Greedy-LB is tighter. In summary, Greedy-LB provides a trade-off between the other two and runs fastest.

\vspace{-0.1in}
\begin{table} [h]
\begin{tabular}
{ l | r | c  c  c c }
\hline
\multirow{2}{*}{} & \multirow{2}{*}{} & \multicolumn{4}{c}{Number of Search State} \\
\cline{3-6}
& & $k=5$ & $k=10$ & $k=15 $& $k=20$ \\
\hline
\multirow{3}{*}{\begin{minipage}{0.38in}DBpedia\end{minipage}} 
& \textbf{Naive-LB}  & 566.9 & 2374.7 & 8973.0 & 15243.2 \\
& \textbf{KM-LB}     & 176.4 & 524.2  & 942.6  & 1737.0 \\
& \textbf{Greedy-LB} & 214.4 & 620.9  & 1062.9 & 2098.3 \\
\hline
\multirow{3}{*}{\begin{minipage}{0.38in}Freebase\end{minipage}} 
& \textbf{Naive-LB}  & 847.8 & 3730.5 & 13371.4 & 21689.9 \\
& \textbf{KM-LB}     & 276.5 & 867.0  & 1535.6  & 2318.2 \\
& \textbf{Greedy-LB} & 284.9 & 1175.2  & 2376.1 & 3709.2 \\
\hline
\end{tabular}
\caption{Evaluating Pruning Power of Lower Bounds. }
\label{tab:search_state_cnt}
\vspace{-0.4in}
\end{table}


\begin{table} [h]
\scalebox{0.92}
{

\begin{tabular}
{ l | r | c  c}
\hline
\multirow{2}{*}{} & \multirow{2}{*}{} & \multicolumn{2}{c}{Avg. Time Cost (ms)} \\
\cline{3-4}
& & DBpedia & Freebase \\
\hline
\multirow{3}{*}{Phase-I} 
& RDF Item Mapping               & 523.6 & 867.0 \\
& Build Candidate Term Graph     & 83.5  & 64.2 \\
& Find Top-$N$ Maximal Clique    & 11.4  & 10.9 \\
\hline
\multirow{3}{*}{Phase-II} 
& Build Assembly Bipartite Graph & 27.6  & 35.1 \\
& Run QGA Algorithm       & 29.1  & 78.5 \\
& Execute SPARQL Query           & 34.7  & 26.7\\
\hline
& \textbf{Overall Response Time} & 734.9 & 1103.2\\

\hline
\end{tabular}
}
\caption{Average Time Cost of Each Step. }
\label{tab:time_cost_each_step}
\vspace{-0.4in}
\end{table}

\subsubsection{Time Cost of Each Step}
Generally, our system's two-phase framework can be further divided into six detailed steps.
We report each step's time cost in Table \ref{tab:time_cost_each_step}. The most time-consuming step is KB Item Mapping, since it involves in expensive I/O operations.
Because Freebase contains more entities, classes and relations than DBpedia, it spends even more time on KB Item Mapping.
Although QGA is NP-complete, our algorithm gains high-efficiency in practice (less than 80ms per query).
We use an open-sourced SPARQL query engine gStore \footnote{https://github.com/Caesar11/gStore.git} \cite{zou2014gstore} to execute the generated SPARQL queries.
Note that any other SPARQL query engine can be used here, which is orthogonal to our task in this paper.

\begin{figure} [t]
	\newcommand{\mywidth}{0.48\textwidth}
	\centering
	\begin{subfigure}[t]{\mywidth}
		\centering
		\scalebox{0.50}
		{
			\includegraphics{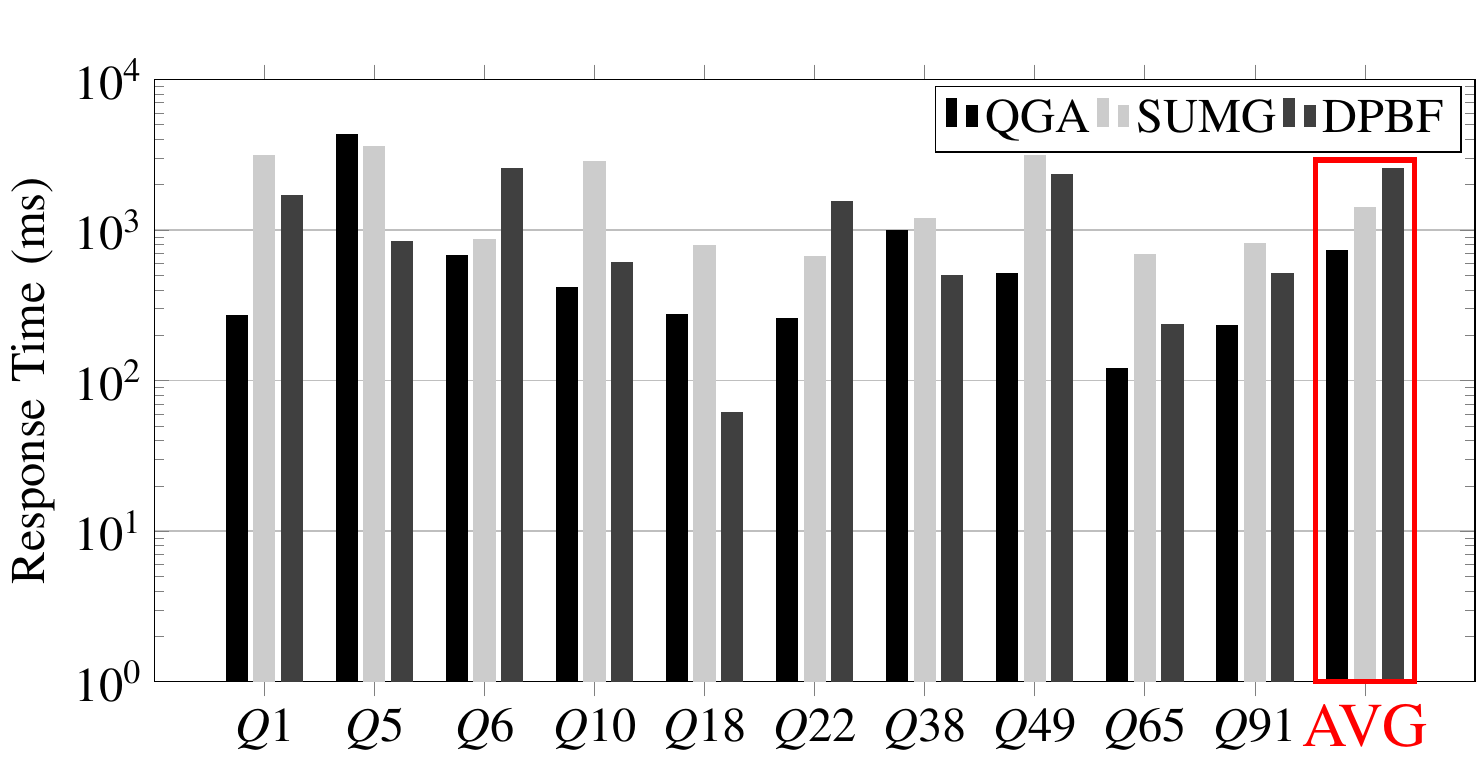}
			
		}
		\vspace{-0.1in}
		\caption{DBpedia + QALD-6}
		\label{fig:timecompare_db}
		\vspace{-0.1in}
	\end{subfigure}
	\begin{subfigure}[t]{\mywidth}
		\centering
		\scalebox{0.50}
		{
			\includegraphics{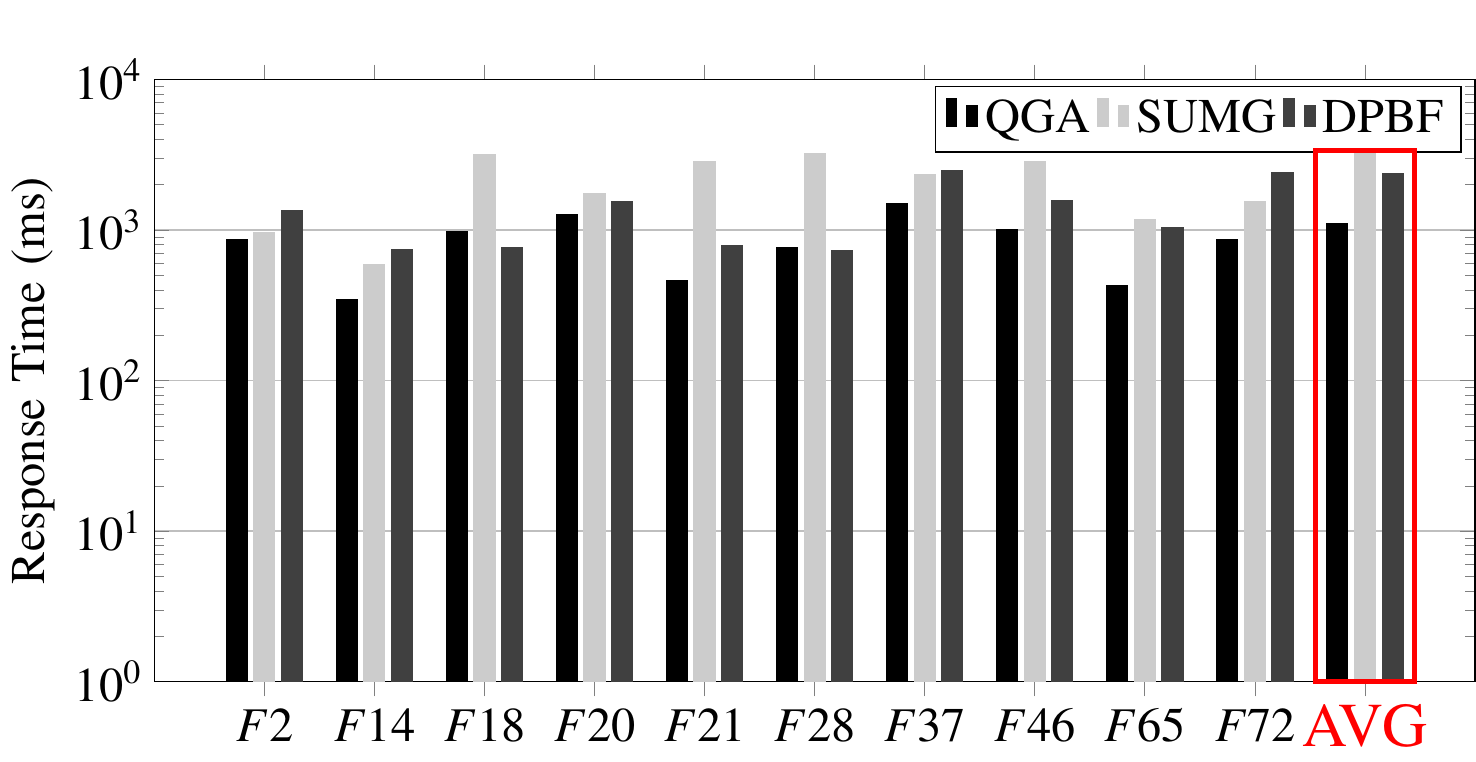}
		}
		\vspace{-0.1in}
		\caption{Freebase + Free917}
		\label{fig:timecompare_fb}
	\end{subfigure}
	\vspace{-0.15in}
	\caption{Overall Response Time Comparison}    
	\label{fig:timecompare}
	\vspace{-0.2in}
\end{figure}

\vspace{-0.05in}
\subsubsection{Comparing with DPBF and SUMG}
We also compare the overall response time.
In Figure \ref{fig:timecompare}, we list the overall response time of 
10 typical queries from QALD-6 and Free917, respectively, and the average response time over all the queries.
Although DPBF and SUMG perform better than our approach on some special cases, such as $Q5$ and $F18$, our approach runs faster than DPBF and SUMG by 1\textasciitilde3 times on average (for all 180 benchmarking queries).

%


\vspace{-0.05in}
\section{Conclusions}
In this paper, we propose a two-phase framework to interpret keyword queries. Different from existing work, we model the keyword search task as assembling a query graph $Q$ to express the query intention. In our solution, the two challenge issues---keyword mapping disambiguation and query structure disambiguation are integrated into one model. We adopt the graph embedding technique (TransE) to measure the goodness of query graphs, which improves the accuracy. Although QGA is proved to be NP-complete, we propose a practical efficient and scalable search algorithm.

\vspace{-0.05in}
\begin{acks}
Lei Zou was supported by the National Key Research and Development Program of China (2016YFB1000603)  and NSFC (No. 61622201 and 61532010).
Jeffery Xu Yu was supported by the Research Grant Council of Hong Kong SAR, China (No. 14221716).
Lei Zou is the corresponding author of this work.
\end{acks}
\vspace{-0.05in}
\bibliographystyle{ACM-Reference-Format}
\bibliography{sigproc} 

\end{document}